\documentclass[final,conference,a4paper]{IEEEtran}

\usepackage[T1]{fontenc}
\usepackage{cite}
\usepackage{url}
\usepackage{amsmath}
\usepackage{amssymb}
\usepackage[cmintegrals]{newtxmath}
\usepackage{subcaption}
\usepackage{dblfloatfix}
\usepackage{graphicx}
\usepackage{epstopdf}
\usepackage{xcolor}
\usepackage{mathtools}
\usepackage[detect-weight=true, binary-units=true]{siunitx}
\usepackage[inline]{enumitem}

\begin{document}
\bstctlcite{IEEEexample:BSTcontrol}

\title{Grouping-Based Random Access Collision Control for Massive Machine-Type Communication}


\author{\IEEEauthorblockN{Bin~Han, Hans~D.~Schotten}
	\IEEEauthorblockA{Technische Universit\"at Kaiserslautern\\
		Institute for Wireless Communication and Navigation\\
		Paul-Ehrlich-Stra\ss e 11, 67663 Kaiserslautern, Germany\\
		Email: \{binhan, schotten\}@eit.uni-kl.de}
}

%
%


\maketitle

\begin{abstract}
	Massive Machine-Type Communication (mMTC) is expected to be strongly supported by future 5G wireless networks. Its deployment, however, is seriously challenged by the high risk of random access (RA) collision. A popular concept is to reduce RA collisions by clustering mMTC devices into groups, and to connect group members with device-to-device (D2D) links. However, analytical models of this method and discussions about the reliability of D2D links are still absent. In this paper, existing grouping-based solutions are reviewed, an analytical model of grouped RA collision is proposed. Based on the analytical model, the impact of D2D reliability on the collision rate is also investigated. Afterwards, an efficient grouped RA procedure is designed to extend the state-of-the-art with an efficient local group update mechanism against D2D link exceptions.
\end{abstract}


\IEEEpeerreviewmaketitle

\section{Introduction}\label{sec:introduction}
Future wireless communication networks of the $5^\textrm{th}$ generation are expected to provide an ultra high traffic density and to connect a huge number of devices. Support to massive machine-type communication (mMTC) is considered as an indispensable part of them. For instance, in the blueprint proposed by the 5G Private Public Partnership (5G PPP), the 5G network shall be capable to connect up to trillions of "things" by 2020\cite{pirinen2014brief,soldani2015horizon}.
Various technical challenges can be caused by such a large amount of MTC devices (MTCDs). The most important one among them is a high risk of random access (RA) collision. Especially, in many MTC scenarios such as sensor networks, MTCDs can be set to transmit or receive data periodically, synchronized with each other. This can lead to instant bursts of RA requests, deteriorating the RA collision problem.

Since the standardization of Long-Term-Everlution (LTE), different methods have been proposed to control the RA collision rate in mMTC, including access class barring (ACB), physical random access channel (PRACH) resource separation, slotted access, dynamic RA resource allocation and MTCD grouping\cite{hasan2013random}. Among them, grouping-based approaches have advantage in resolving collisions in duty-cycle-critical scenarios \cite{wang2013random}, and has attracted much interests in recent years. 
Especially, many grouping-based approaches invoke device-to-device (D2D) connections for the intra-group communications \cite{chatzikokolakis2015way,chuang2015group}, and present satisfying result in reducing RA collision rates. However, these proposed methods have only been evaluated through simulations under specific system configurations, while analytical models for the efficiency are absent. Besides, they generally consider the D2D links as available and reliable, ignoring the risk of D2D connection failures. This assumption, however, can unfortunately deviate from the practical truth. In this paper, we analytically investigate the RA collisions in grouped mMTC scenarios, and discuss about the impact of D2D connection reliability on grouping-based RAN congestion control. We also extend the state of the art with an advanced protocol, which aims at reducing RA traffic with an efficient local group update mechanism.

The paper is structured as follows. In Sec. \ref{sec:grouping} we introduce the concept of MTCD grouping, review existing approaches, especially D2D-based solutions, and point out the open issues to study. In Sec.\ref{sec:analysis} we propose analytical models of the collision rate and collision intensity in grouped RA procedures, and discuss about the impacts of group size and D2D link failures. In Sec. \ref{sec:procedure} we present our novel grouped RA protocol in details. At the end we close this paper with some conclusion and outlooks in Sec. \ref{sec:conclusion}.

\section{Grouping-Based RAN Congestion Control}\label{sec:grouping}

\subsection{Current Grouping-Based Solutions}
Generally, all RAN congestion controlling methods that we call as grouping-based try to cluster UEs into several groups. The idea of grouping UEs had been initially proposed to raise battery life\cite{kim2010snoop}, but later on it was also recognized to possess an advantage in avoiding RAN congestions\cite{lien2011toward,tu2011energy}. By assigning one group coordinator (GC) to each group and letting the GC sending RA preambles for its entire group, the total amount of preamble sources in the cell is decreased, hence the congestion rate can be reduced. 

To guarantee a high utilization efficiency of RAN resource, only the UEs that truly need connections should be granted uplinks. Therefore, when receiving a preamble from a GC, the base station (BS) has to know which UEs in this group are requesting connections. It would be one straight-forward solution, that the GC collects RAN requests from its group members (GMs), and provides the BS a list of GMs requesting connections. However, this leads to huge signaling overheads, not only between the GMs and the GC, but also between the GC and the BS. Especially for the second one, a long list of GM identities must be transmitted to the BS. It is impossible to embed so much information into the preambles in their current formats. Hence, either a new format must be designed for the preambles, or an additional message must be sent to the BS in the RA procedure. Both methods will make it difficult to keep a seamless backward compatibility to existing LTE/LTE-A systems. Therefore, an alternative solution has been more deployed: to make sure that all GCs in every group always request connections together.

This solution is enabled by the fact that many mMTC devices are designed to work in a synchronous mode, i.e. they transmit / receive data with a same time period. Hence, if the groups are constructed according to the device type, ensuring that all the devices in each group have the same period of data transmission, the BS can simply grant radio resource control (RRC) link to all the GMs, once a preamble is received from their GC. Schemes of this kind have been proposed in \cite{chatzikokolakis2015way} and \cite{lee2012group}, for instances. 

The assignment of RRC links is also often of interest in grouping-based schemes. Generally, three classes of solutions are available. First, individual RRC links can be granted to different GMs. In this case, each GM must receive a unique RA responses from the BS, which can be achieved with techniques such as slotted access\cite{lee2012group}. Second, all GMs in a same group can share the same RRC link in multiplex, which may bring in extra signaling overhead and is therefore not preferred. To the third, device-to-device (D2D) connections can be deployed to help the GC work as a gateway, aggregate uplink data from its GMs or distribute downlink data to them \cite{chatzikokolakis2015way,chuang2015group}. Compared to the first method, the D2D-based solution is encouraged by its lower power consumption, lower RACH overhead and higher RAN resource efficiency. 
Hence, in the following part we focus on this method, taking the method proposed in \cite{chatzikokolakis2015way} as example for introduction and discussion.

\subsection{Grouped RA Scheme Based on D2D Communication}

In \cite{chatzikokolakis2015way}, a hierarchical device classification method is designed, where every MTCD uploads its information to the BS during first attachment, including transmission periodicity, data size, packet delay, device mobility, etc. According to the information, every group is supposed to contain only MTCDs with the same transmission period and similar packet delay. Asynchronous devices are never clustered into any group. 

\begin{figure}[!h]
	\centering
	\begin{subfigure}{0.49\textwidth}
		\centering
		\includegraphics[width=.7\textwidth]{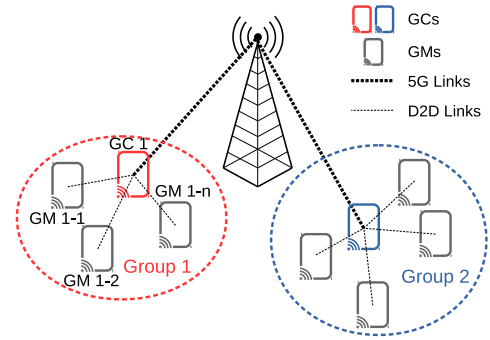}
		\caption{In a clustered mMTC network, only group coordinators periodically directly communicate with the BS via 5G links, while other devices only communicate with their own group coordinators via D2D links.}
	\end{subfigure}
	\begin{subfigure}{0.49\textwidth}
		\centering
		\includegraphics[width=.5\textwidth]{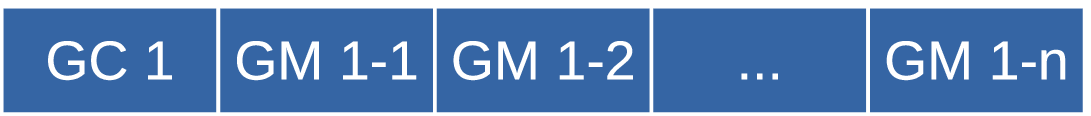}
		\caption{Every user data package transmitted between GC and BS contains several segments, each segment for a particular device.}
		\label{fig:aggregated_user_data}
	\end{subfigure}
	\begin{subfigure}{0.49\textwidth}
		\centering
		\includegraphics[width=.6\textwidth]{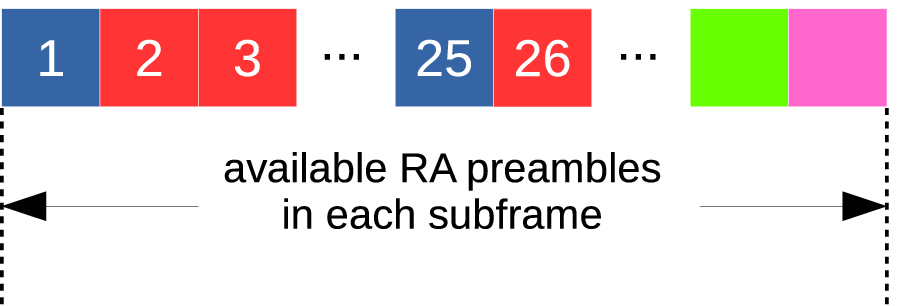}
		\caption{RA preambles available for mMTC can be allocated to different device types, which are represented in different colors.}
	\end{subfigure}
	\begin{subfigure}{0.49\textwidth}
		\centering
		\includegraphics[width=.6\textwidth]{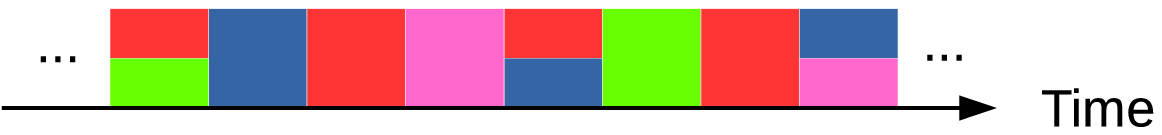}
		\caption{Through a time scheduling process, each RACH time slot can be reserved for one or multiple perticular device types.}
	\end{subfigure}
	\caption{The grouped RA scheme originally proposed in \cite{chatzikokolakis2015way}}
	\label{fig:grouped_ra_scheme}
\end{figure}

Afterwards, a time-scheduling method is proposed, as illustrated in Fig. \ref{fig:grouped_ra_scheme}. In each RACH time frame, each group is addressed to several RA preambles and several time slots, according to its device type. Each preamble or time slot can be either dedicated to one group, or shared by multiple groups of different device types. When a GC requests connections for its group, it randomly selects one from its available RA preambles, and send it to the BS when the first addressed slot comes. If no congestion occurs, the BS then sends its response to the GC, granting it an uplink channel, so that an RRC connection can be established between the BS and the GC, which serves all UEs in the group. User data from / for all UEs are first aggregated and packed with a header segment to declare the packet structure, before it is transmitted. According to the header information, the BS / GC can extract the UL / DL data packet into segments from / for different UEs, respectively. The intra-group aggregation and distribution of user data are accomplished over D2D links.

According to simulation results\cite{chatzikokolakis2015way}, this approach shows a satisfying performance of reducing the collision rate by around $75\%$. However, some important issues still remain open. 

\subsection{Open Issues}\label{subsec:open_issues}

First, most existing D2D-based solutions have been only evaluated through network simulations, without giving an analytical performance model, which can be beneficial for exploration of their dependencies, their sensitivities to system and environment factors, as well as their performance limits.

Next, all D2D-based solutions highly rely on the stability of D2D connections between GC and GMs. In practice, depending on the geo-locations of UEs and the electro-magnetic (EM) environment, a GM may fail to build a stable D2D connection to its assigned GC, or lose the established D2D connection, despite its availability to the BS. In these cases, it will fail to access the BS via its GC. Reporting or acknowledging mechanism can be designed in the protocol to handle such exception, however, it may create extra overhead and thus reduce the efficiency of congestion control. Up to now, this impact has not been studied yet.

Furthermore, due to the mobility and dynamics of UEs, MTCD groups have to be dynamically updated. A cell-wide update of groups is essential to support initialization and optimization,  but also generates a large overhead. Hence, it should only be triggered in a low frequency. Therefore, supplementary processes are needed to update a certain MTCD group, in case a single device joins or leaves it. Detailed message flows for these processes have not been reported yet.

In the sections below, we discuss these issues one by one.

\section{Collision Analysis in Grouped Random Access}\label{sec:analysis}
\subsection{Random Access Collision Rate in LTE}
An analytical study on the RA collision probability for mMTC was presented in \cite{3gpp2011study}. When a MTCD sends a preamble to the BS, the probability that it causes a collision is
\begin{equation}\label{equ:collision_rate_basic}
P_\textrm{c} = 1-e^{-\frac{\gamma}{L}},
\end{equation}
where $\gamma$ and $L$ denote the average RA request intensity per second and $L$ the total number of available random access opportunity (RAO) per second, respectively.

Classifying the MTCDs into an asynchronized class and $M$ different synchronized classes, we can rewrite (\ref{equ:collision_rate_basic}) as
\begin{equation}
P_\textrm{c} = 1-e^{-\sum\limits_{i=0}^M\frac{\gamma_i}{L}},
\end{equation}
where $\gamma_i$ is the average intensity of RA requests made by MTCD of the $i^\textrm{th}$ class ($i=0$ for asynchronized devices and $1\le i\le M$ for periodic devices).
Here, it worths note that $\gamma$ can be significantly time-varying in highly synchronized mMTC scenarios, further raising the peak collision risk. In the optimal case, where all RA request arrivals are uniformly distributed through time, there is
\begin{equation}\label{equ:lte_collision_rate_optimal}
\gamma_i=\frac{N_i}{T_i}\qquad i=1,2,\dots,M
\end{equation}
where $N_i$ and $T_i$ are the number of MTCDs of the $i^\textrm{th}$ device class and its period, respectively. In the worst case, when all synchronized devices send RA requests at the same time, e.g. the first second of every day, the instant value of $\gamma_i$ reaches the peak at this moment:
\begin{equation}\label{equ:lte_collision_rate_worst}
\gamma_{i,\max}=N_i\qquad i=1,2,\dots,M
\end{equation}
Note that (\ref{equ:lte_collision_rate_optimal}) and (\ref{equ:lte_collision_rate_worst}) are invalid for aperiodic devices ($i=0$).

\subsection{Collision Rate in Grouped Random Access Procedure}
Grouping-based RAN congestion control approaches reduce $P_\textrm{c}$ by reducing $\gamma$. First, as only GCs will send RA preambles, (\ref{equ:lte_collision_rate_optimal}) turns into
\begin{equation}\label{equ:periodic_grouped_arrival}
\tilde{\gamma}_i=\frac{n_i}{T_i}\qquad i=1,2,\dots,M,
\end{equation}
where $n_i$ is the number of groups of the $i^\textrm{th}$ device class (DC). 
For convenience of expression we also define
\begin{equation}\label{equ:asyn_ra_no_d2d_exception}
\tilde{\gamma}_0=\gamma_0.
\end{equation}
For simplification here we assume that each class contains $K$ devices (except for the residues), so that
\begin{equation}\label{equ:periodic_grouped_arrival_constant_group_size}
n_i = \lceil \frac{N_i}{K}\rceil\qquad i=1,2,\dots,M.
\end{equation}
Then, through an appropriate time frame scheduling, groups can be configured to send their RA requests in an optimal time distribution, i.e. uniformly distributed through time. Thus, the RA request arrival intensity always remains on its average level, and burst cases such as (\ref{equ:lte_collision_rate_worst}) can be avoided.

As introduced before, each preamble can be either dedicated to one specific DC, or shared by multiple DCs, which leads to a more complex calculation of the collision rate. When a MTCD of class $i$ sends a preamble $l$, the collision rate is
\begin{equation}
P_{\textrm{c},i,l}=1-e^{-\sum\limits_{j\in\mathbb{A}_l}\frac{\tilde{\gamma}_j}{L_j}}
\end{equation}
where $\mathbb{A}_l$ is the set of all DCs that can use the preamble $l$ and $L_j$ is the total number of RAO for the DC $j$. As the device randomly selects the preamble to send, the expectation of collision rate when it sends an arbitrary preamble is
\begin{equation}\label{equ:preamble_allocation}
P_{\textrm{c},i}=\frac{1}{\sharp\mathbb{B}_i}\sum\limits_{l\in\mathbb{B}_i}P_{\textrm{c},i,l},
\end{equation}
where $\mathbb{B}_i$ is the set of all preambles available for the DC $i$.

Two extreme cases of the RAO allocation strategy are full RAO sharing and full RAO dedication. In the first case, all RAOs are available for all DCs, so that $L_j=L$ for all $j$, $\mathbb{A}_l=\mathbb{A}=\{0,1,\dots,M\}$ for all $l$ and $\mathbb{B}_i=\mathbb{B}$ for all $i$, where $\mathbb{B}$ is the complete set of all preambles. Hence, (\ref{equ:preamble_allocation}) turns into
\begin{align}\label{equ:collision_rate_sharing}
P_{\textrm{c},i,\textrm{sharing}}&\triangleq P_{\textrm{s},i}=\frac{1}{\sharp\mathbb{B}}\sum\limits_{l\in\mathbb{B}}(1-e^{-\sum\limits_{j\in\mathbb{A}}\frac{\tilde{\gamma}_j}{L}})\nonumber\\
&=1-e^{-\sum\limits_{j=0}^M\frac{\tilde{\gamma}_j}{L}}=1-e^{-\sum\limits_{j=1}^M\frac{\lceil N_j/K\rceil}{LT_j}-\frac{\gamma_0}{L}}\nonumber\\
&= P_{\textrm{s}},
\end{align}
which is independent of $i$, showing no DC preference. 

In the second case, each RAO is dedicated to a specific DC. Obviously, this strategy disables any cross-class collision, so we can simplify the grouped RA collision problem by decomposing it into $M+1$ independent ungrouped RA collision problems in the form of (\ref{equ:collision_rate_basic}):
\begin{align}\label{equ:collision_rate_dedication}
P_{\textrm{c},i,\textrm{dedication}}&\triangleq P_{\textrm{d},i}=1-e^{-\frac{\tilde{\gamma}_i}{L_i}}\nonumber\\
&=\begin{cases}
1-e^{-\frac{\gamma_0}{L_0}}\qquad &i=0\\
1-e^{-\frac{\lceil N_i/K\rceil}{L_iT_i}}\qquad &i=1,2,\dots,M
\end{cases},
\end{align}
which is a function of $i$ due to its dependency on $\frac{\tilde{\gamma}_i}{L_i}$. 


Equations (\ref{equ:preamble_allocation})--(\ref{equ:collision_rate_dedication}) mathematically support our intuitive idea, that the correlation between the collision rates of different DCs increases if they share more RAOs, and vice versa. 

Moreover, the overall collision intensity in a cell can be calculated by
\begin{align}\label{equ:grouped_collision_intensity}
&C=\sum\limits_{i=0}^{M}P_{\textrm{c},i}\tilde{\gamma}_i\nonumber\\=&\begin{cases}
(1-e^{-\sum\limits_{j=0}^{M}\frac{\tilde{\gamma}_j}{L}})\sum\limits_{i=0}^M\tilde{\gamma}_i&\textrm{full RAO sharing}\\
\sum\limits_{i=0}^M(1-e^{-\frac{\tilde{\gamma}_i}{L_i}})\tilde{\gamma}_i&\textrm{full RAO dedication}\\
\sum\limits_{i=0}^M\tilde{\gamma}_i\frac{1}{\sharp\mathbb{B}_i}\sum\limits_{l\in\mathbb{B}_I}(1-e^{-\sum\limits_{j\in\mathbb{A}_l}\frac{\tilde{\gamma}_j}{L_j}})&\textrm{otherwise}
\end{cases}.
\end{align}


\subsection{Impact and Constrains of Group Size}
Clearly, by increasing the group size $K$ we can decrease $n_i$ and therefore also $\gamma_i$ for all $1\le i\le M$, as (\ref{equ:periodic_grouped_arrival}) and (\ref{equ:periodic_grouped_arrival_constant_group_size}) show:
\begin{equation}
\frac{\partial\tilde{\gamma}_i}{\partial K}<0\qquad i=1,2,\dots,M.
\end{equation}
According to ($\ref{equ:grouped_collision_intensity}$), this leads to a decreased collision intensity under all RAO allocation strategies, when the asynchronous RA request intensity $\tilde{\gamma}_0$ remains independent of $K$. Therefore, a large $K$ is generally expected. The ideal case will be, then, when $K$ is large enough so that for every periodic DC there is only one group, i.e. $n_i=1$ and $\tilde{\gamma}_i=\frac{1}{T_i}$ for all $1\le i\le M$. However, this is usually impossible due to multiple implementation constrains.

First, $K$ is limited by the size of aggregated user data package. As shown in Fig. \ref{fig:aggregated_user_data}, the length of each data package transmitted between a GC and the BS is sum of $K$ segments. The maximal length of a data package is limited by the RACH time slot specification, while the length of each segment is determined by the MTC message content. Hence, there is an upper bound of $K$. Second, common D2D technologies mostly support link range significantly shorter than the macro cell size. Unless all MTCDs were spatially clustered according to their device classes (DCs), which is unrealistic, all devices of same DC in a macro cell cannot be covered by a single group.

Moreover, a large MTCD group size $K$ can also lead to drawbacks, even when it does not exceed its upper bound limited by these hard constrains. For example, the overall throughput of a GC, including both 5G and D2D traffics, is proportional to the size of its group. A large $K$ will cause high power consumption of the GC, significantly shortening its battery life and reducing the intra-group energy fairness. Periodical GC re-selection can be applied to overcome this problem, but still leaving a high peak-to-average ratio of power consumption on every device, which is not preferred. More interestingly, when taking the D2D link stability into our consideration, as we will see in the next section, the independence of $\tilde{\gamma}_0$ on $K$ is no more guaranteed. Increasing the group size may create extra asynchronous RA requests, which compensate the gain from decreased periodic RA requests.

\section{Impact of D2D Link Exceptions on Random Access Collision Rate}
Existing studies on D2D-based grouping solutions for RAN congestion control generally consider D2D links as available and reliable, as long as the 5G service is also available. However, this assumption is not guaranteed to be true. Since many D2D technologies use unlicensed spectrum, narrowband noises can interfere the D2D connections without influencing the 5G connection. Besides, shadowing objects such as buildings can cause direction-selective attenuations. Hence, when a MTCD is clustered into some group as a GM, it may fail to establish a D2D connection with its dedicated GC. Furthermore, due to the user mobility and the environment dynamics, the path loss and interferences can be usually time-varying, so that even established D2D connections can also vanish or collapse. When such events happen, the BS must be informed, in order to update the grouping map. As the GM is proposed to communicate with the GC instead of directly with the BS, there is no stable uplink channel promoted to it. An extra RA procedure between the MTCD and the BS is therefore needed, to accomplish such an exception report. The D2D link exceptions arraival randomly and asynchronously, so the extra RA procedures will generate asynchronous RA requests, and hence increase $\tilde{\gamma}_0$.

Given a certain set of MTCDs in the same DC and in the same local cell, we tend to cluster them into several groups. It is reasonable to assume, that the arrival rate of D2D link exceptions decreases when we have more and smaller groups, and increases when we have less but larger groups, because:
\begin{itemize}
	\item Larger groups must have large radius of coverage, leading to a higher average GC-GM distance, and hence a lower SNR of D2D connections.
	\item Larger group size means fewer GC in the entire cell, and more D2D connections per GC, resulting in a higher probability that a link is shadowed by terrains or buildings.
\end{itemize}
In short words, there is
\begin{equation}
\frac{\partial \tilde{\gamma}_0}{\partial K}>0,
\end{equation}
which breaks the monotonicity of $C$ on $K$.

\begin{figure}[!h]
	\centering
	\includegraphics[width=.4\textwidth]{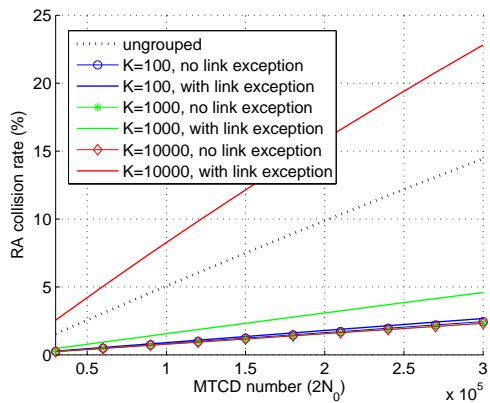}
	\caption{Grouped RA collision rate under the full RAO sharing strategy, with consideration of D2D link exception reports. Ungrouped RA collision rate and reliable D2D links are also presented for comparison.}
	\label{fig:grouped_ra_collision_rate}
\end{figure}

For a brief demonstration, we assumed that the exception intensity of each D2D link is an exponential function of the group size, and replace (\ref{equ:asyn_ra_no_d2d_exception}) with a simple model:
\begin{equation}
\tilde{\gamma}_0=\gamma_0+(e^{\alpha (K-1)}-1)\sum\limits_{i=1}^MN_i,
\end{equation}
where $\gamma_0$ is the intensity of normal asynchronous RA requests, $e^{\alpha (K-1)}-1$ is the expectation of D2D link exception amount that an arbitrary MTCD experiences per second, and $\alpha$ is a parameter for the sensibility of exception rate on $K$. The model was designed to let $\tilde{\gamma}_0=\gamma_0$ when $K=1$, because there is no D2D link to establish in this case. Then we took the device class specifications used for simulation in \cite{chatzikokolakis2015way}, where $M=6$ different periodical DCs are available, as listed in Tab. \ref{table:simu_config}. We also set $L=3600$, $\gamma_0=0.001$, and $\alpha=1\times10^{-6}$, which approximately corresponds to a D2D link exception rate of $1\times10^{-3}\si{\hertz}$ when $K=1000$. Thus, we calculated the RA collision rates under the full RAO sharing strategy according to (\ref{equ:collision_rate_sharing}), as shown in Fig. \ref{fig:grouped_ra_collision_rate}. When not considering the link exceptions, a gain very similar to the results reported in \cite{chatzikokolakis2015way} can be achieved, and the gain slightly increases, when the group size grows. However, when taking the extra RA traffic caused by link exceptions into account, this gain can be significantly reduced, or even overwhelmed when the groups are huge.
\begin{table}[!h]
	\caption{Specifications of DCs for the simulation in \cite{chatzikokolakis2015way}}
	\label{table:simu_config}
	\centering
	\begin{tabular}{c|c|c|c|c|c|c|c}
		Device class & 0 & 1 & 2 & 3 & 4 & 5 & 6\\\hline
		Periodicity & N/A & \SI{1}{\minute} & \SI{1}{\minute} & \SI{1}{\hour} & \SI{1}{\hour} & \SI{1}{\day} & \SI{1}{\day}\\\hline
		Number & $N_0$  & $N_0/6$ & $N_0/6$ & $N_0/6$ & $N_0/6$ & $N_0/6$ & $N_0/6$
	\end{tabular}
\end{table}

Therefore, in grouping-based mMTC RA collision control, a reasonable group size $K$ must be carefully selected with respect to the D2D link reliability.

Besides, as aforementioned to in Sec. \ref{subsec:open_issues}, a global update of all MTCD groups leads to a huge burst of RA traffic, as it requires direct communication between the BS and each single MTCD. Hence, to reduce $\tilde{\gamma}_0$, the grouped RA procedure must support local group updates, such as adding/removing a single MTCD to/from a group, without affecting other groups. In the next section, our procedure design will be introduced.

\section{Grouped Random Access Protocol with D2D Link Exception Handling Mechanism}\label{sec:procedure}
To construct MTCD groups and update them efficiently, three kinds of operations are essential in the grouped RA procedure: global group update, group joining and group leaving. With respect to the triggering event, the message flow of each operation can vary in different cases.

\subsection{Global Group Update}
Our discussion begins with the global group update, which can be triggered by three different events:
\begin{enumerate}
	\item \textit{Initial clustering}: no group is available in the cell yet, and the RAN congestion rate exceeds a given threshold. MTCDs shall be therefore clustered.
	\item \textit{Emergency re-clustering}: groups are already available in the cell, the RAN congestion rate exceeds a given threshold. Groups must be therefore optimized.
	\item \textit{Regular re-clustering}: groups are already available in the cell, a certain timeout has past since the last global group update. Groups shall be updated to remain optimal.
\end{enumerate}

\begin{figure}[!h]
	\centering
	\includegraphics[width=.3\textwidth]{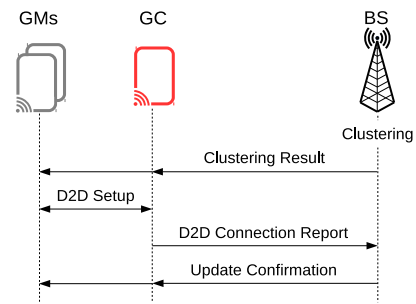}
	\caption{Global group updating}
	\label{fig:mfd_ggu}
\end{figure}

All three triggers can be detected at the BS side, so the procedure begins at BS, as shown in Fig. \ref{fig:mfd_ggu}. BS calls from database the context information of all MTCDs (device type, geo-location, connection quality, etc.), which is periodically updated. Then it executes a clustering algorithm, clusters all periodical MTCDs into several groups according to the following disciplines:
\begin{enumerate}
	\item Every group only contains MTCDs of the same device type, to promise the synchronization.
	\item The geographical diameter of each group is limited to a maximum, to promise the D2D coverage.
	\item The total number of members in each group is limited to a device-type-dependent maximum, to promise an efficient data aggregation / distribution in a time window.
\end{enumerate}
Afterwards, for every group, a GC is selected.

We call the member list and the GC selection of a group as its \textit{clustering result}. The BS then transmits the clustering result to the GC and all GMs of every group. According to this message, every GC releases D2D access to its GMs, and all GMs try to visit the GC. Note again that failures can happen to some devices during the D2D connection setup. Either after successfully setting up all D2D connections, or after a timeout, the GC reports to BS about the setup result. If any GM fails in the setup, it will be removed from the group list on the BS side. At last, the BS transmits the updated clustering result to every group for a confirmation. Successfully grouped devices then turn into their GM/GC mode after receiving the confirmation message, while failed devices keep working in ungrouped mode, until the next group update.

\subsection{Group Joining}
A new device may enter a cell, when a handover happens or when the device tries to initiate a network attachment in the cell. In the former case, the BS is informed about the handover and the device's information by the neighbor BS, which has been serving the device, so that it is able to immediately initiate the group joining process. In the latter case, the process can only be triggered after the first attachment, during which the device uploads its context information such as device type and geo-location to the BS. But generally, in both case the group joining process is initiated by the BS.

We design the group joining process in a very similar way to the global group update, with only two differences:
\begin{enumerate}
	\item The BS does not re-cluster all devices in the cell. Instead it selects the best existing group for the joining device. If no group is appropriate, the process will be terminated.
	\item After selecting the best group, the BS transmits its decision only to the respective GC and the joining device. A D2D setup will be attempted, then the GC reports the result to the BS for an update confirmation. Other GMs of the selected group and the devices in other groups are not involved in this process.
\end{enumerate}

\subsection{Group Leaving}
One clustered device may leave its current group when it:
\begin{enumerate}
	\item requests to detach, e.g. while turning idle or off;
	\item as a GM, unexpectedly loses the connection to its GC but remains able to connect to the BS ;
	\item as a GM, unexpectedly loses the connections to its GC and the BS ;
	\item as a GC, unexpectedly loses the connection to the BS;
	\item moves from the current cell towards another (handover).
\end{enumerate}

\begin{figure}[t]
	\centering
	\begin{subfigure}{0.49\textwidth}
		\centering
		\includegraphics[width=.8\textwidth]{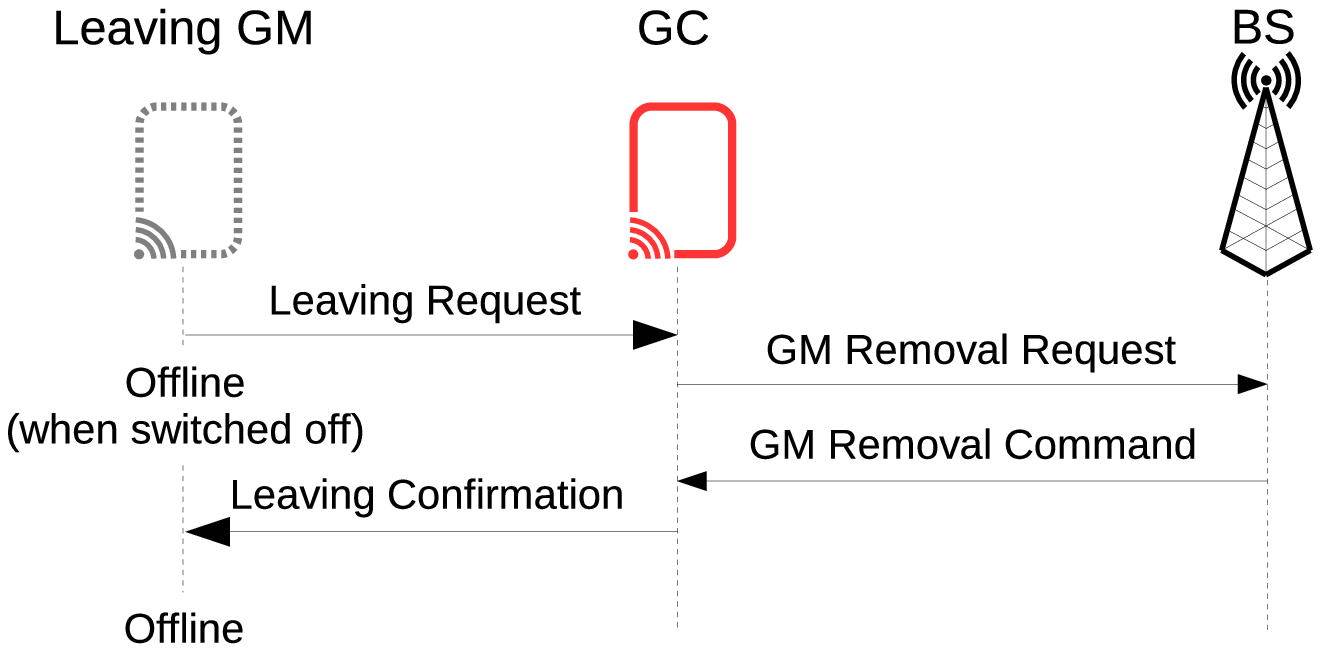}
		\caption{A GM is switched offline.}
		\label{fig:mfd_gl_gm_offline}
	\end{subfigure}
	\begin{subfigure}{0.49\textwidth}
		\centering
		\includegraphics[width=.9\textwidth]{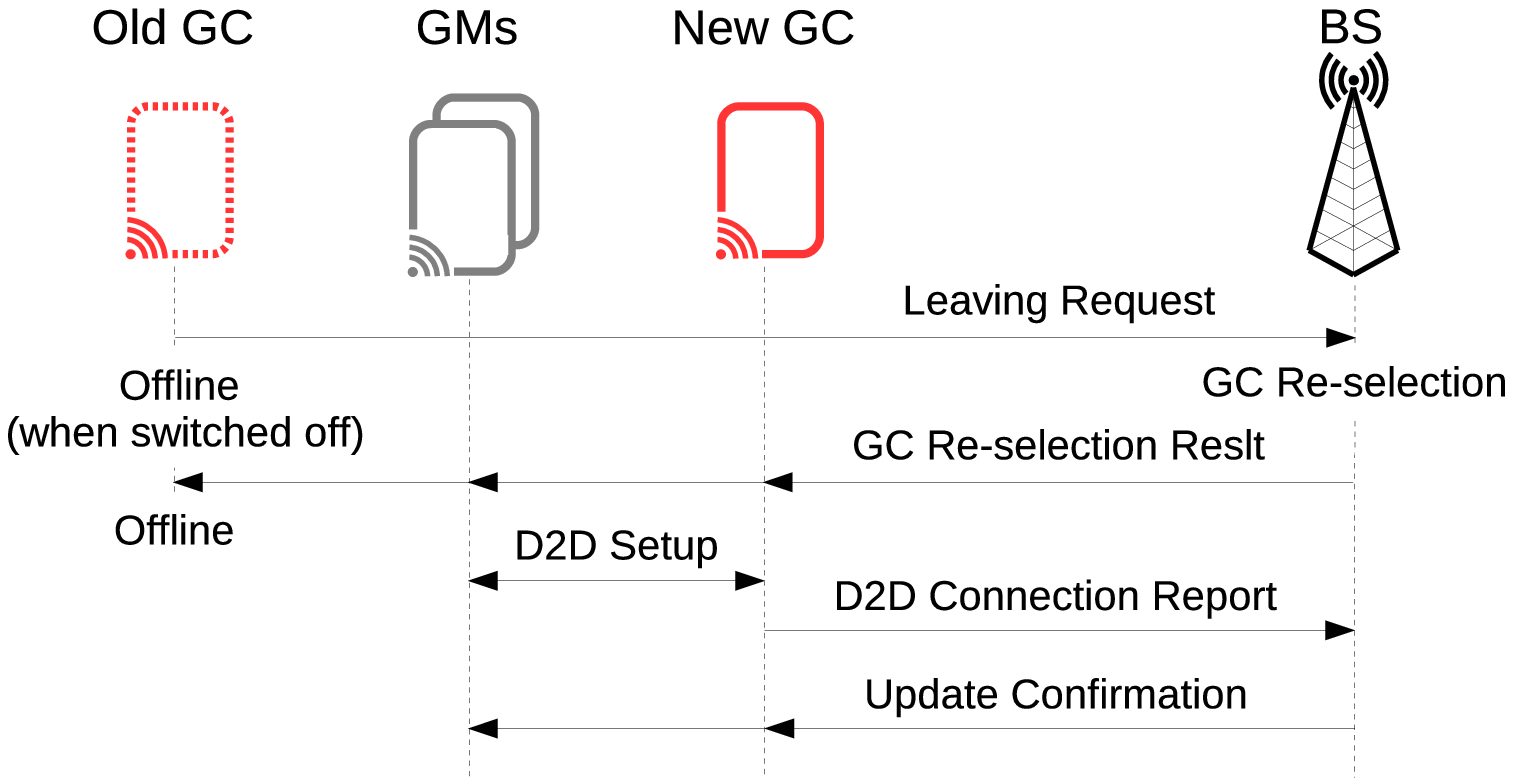}
		\caption{A GC is switched offline.}
		\label{fig:mfd_gl_gc_offline}
	\end{subfigure}
	\caption{Group leaving triggered by switching-off events}
	\label{fig:mfd_gl_offline}
\end{figure}

The proposed group leaving process in the detaching case is illustrated in Fig. \ref{fig:mfd_gl_offline}. Generally, the device initiates this process. Depending on its role in group (GM or GC), two different procedures are applied, as shown in Figs. \ref{fig:mfd_gl_gm_offline} and \ref{fig:mfd_gl_gc_offline}, respectively. If it is a GM, only the BS and GC shall be informed, to remove it from the GM-list. If it is the GC, a new GC must be selected to replace it. If the detaching process is started due to a switching-off operation, the device turns offline immediately after sending the leaving request. Otherwise, it waits for confirmation before going offline.

\begin{figure}
	\centering
	\includegraphics[width=.4\textwidth]{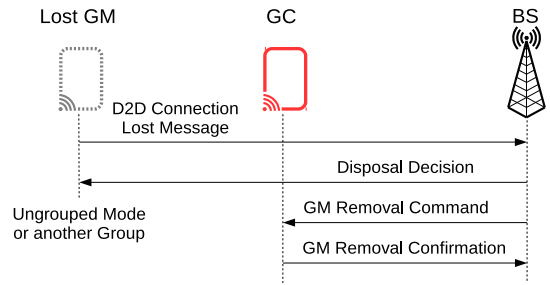}
	\caption{Group leaving triggered by a GM, which unexpectedly loses its D2D connection to GC but can still access BS}
	\label{fig:mfd_gl_gm_lost_d2d}
\end{figure}

When a GM loses its D2D connection to GC but remains available for the BS, the group leaving process is executed as shown in Fig. \ref{fig:mfd_gl_gm_lost_d2d}. After recognizing the disconnection, the GM reports to the BS. Such D2D disconnections can be caused by various reasons, including changed channel condition, device mobility, etc. Hence, the report is supposed to include some context information of the GM, e.g. its current location, noise power measurement, etc. Based on this report, the BS makes a decision of disposal, either letting the GM switch into ungrouped mode, or cluster it into another group (which initiates a group-joining process). Afterwards, the GC should be ensured to know about this operation of GM removal.

\begin{figure}
	\centering
	\includegraphics[width=.3\textwidth]{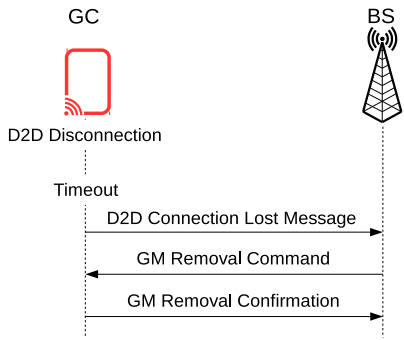}
	\caption{Group leaving triggered by a GM, unexpectedly loses its D2D connection to GC and cannot access BS}
	\label{fig:mfd_gl_gm_lost_5g}
\end{figure}

In a worse case, if the GM loses both its D2D connection and its 5G connection, it cannot report the disconnections by itself. In this case, the GC takes the responsibility to inform the BS about it. To avoid conflict with the last case, the GC should wait for a certain timeout after recognizing the disconnection. Only when no GM removal command is received from the BS during this period, it reports the BS, as shown in Fig. \ref{fig:mfd_gl_gm_lost_5g}.

When a GC unexpectedly disconnects from the BS, the BS recognizes it immediately. In this case, the GC is removed from its group and a new GC is selected, leading to a single-group update, which is similar to the process in Fig. \ref{fig:mfd_ggu}.

\begin{figure}
	\centering
	\begin{subfigure}{0.49\textwidth}
		\centering
		\includegraphics[width=.8\textwidth]{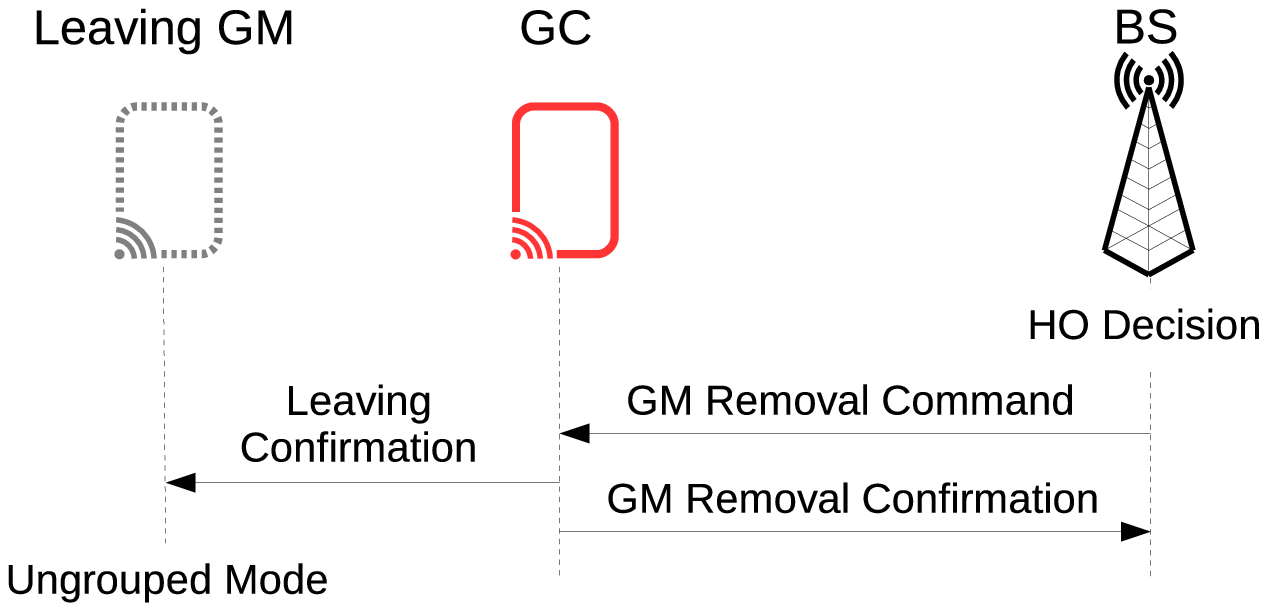}
		\caption{A GM moves towards another cell.}
		\label{fig:mfd_gl_gm_ho}
	\end{subfigure}
	\begin{subfigure}{0.49\textwidth}
		\centering
		\includegraphics[width=.9\textwidth]{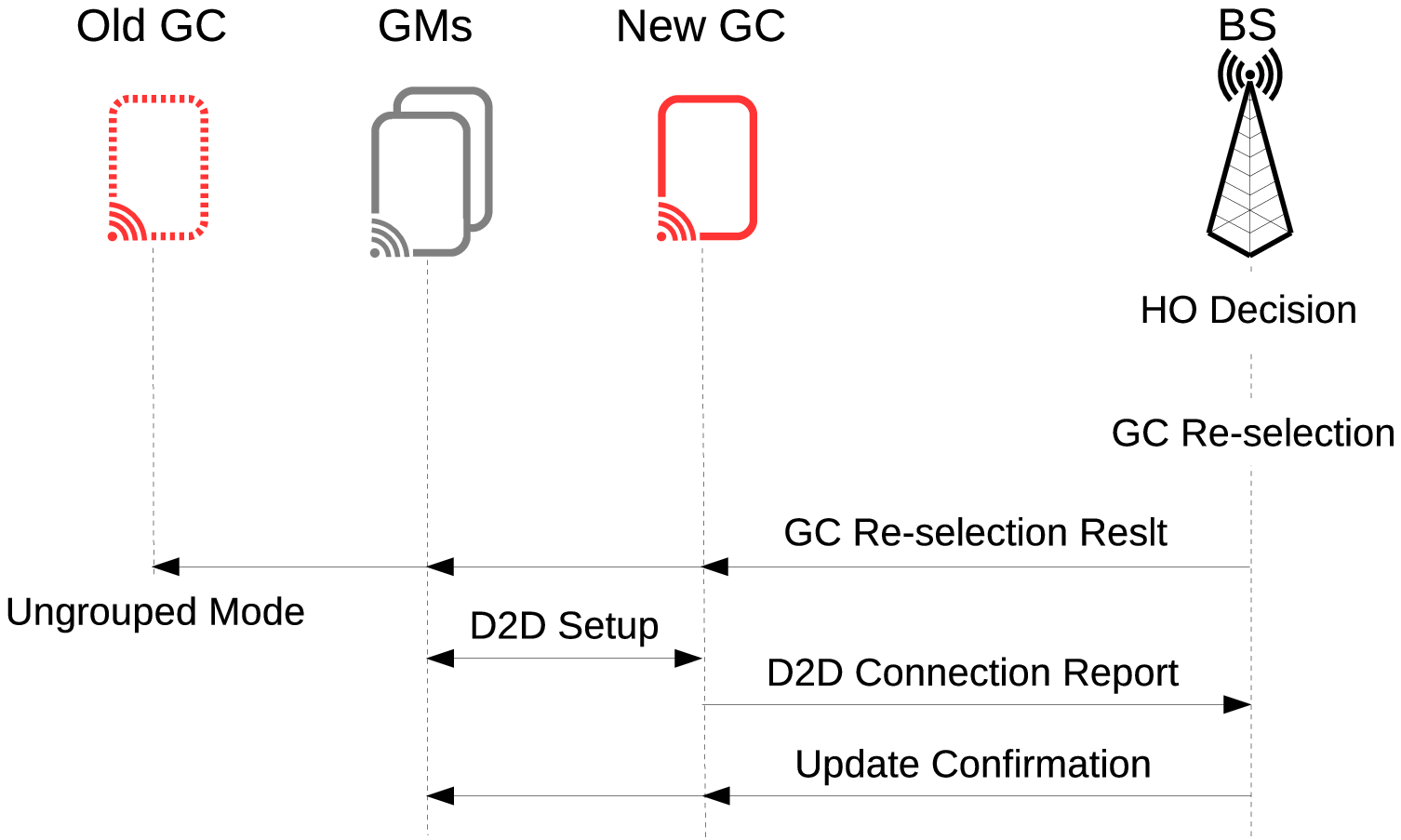}
		\caption{A GC moves towards another cell.}
		\label{fig:mfd_gl_gc_ho}
	\end{subfigure}
	\caption{Group leaving triggered by handover events}
	\label{fig:mfd_gl_ho}
\end{figure}

A handover event indicates a detachment from the source BS and an attachment to the target BS. If the MTCDs in the target cell are clustered, however, is unknown. To simplify the decision chain, we always turn the leaving device into ungrouped mode before the handover process is completed, and let the target BS evaluate the necessity of initiating a group joining process. When the source BS makes a HO decision, it examines the group role of the device in handover. If it is a GM, it is removed from the list of GMs, and the GC is informed about this removal, as Fig. \ref{fig:mfd_gl_gm_ho} shows. If the GC is in handover, a new GC must be selected instead, as depicted in Fig. \ref{fig:mfd_gl_gc_ho}.

\section{Conclusion and Outlooks}\label{sec:conclusion}
So far, we have studied the efficiency of D2D-based MTCD grouping for RA collision control, with respect to RAO allocation strategy, group size and D2D link exception rate. We have analytically showed that increasing the group size helps reduce RA collisions from periodical MTCDs, but the drawback of more D2D link exceptions may overcome the gain obtained from large group sizes. We have also designed an efficient group RA procedure to reduce asynchronous RA requests.

For future study, we look forward to a deeper analysis on the general RAO allocation strategy. A realistic model of D2D exception rate with respect to group size is also to our full interest. Based on these knowledge, a group optimization algorithm can be eventually implemented.

\section*{Acknowledgment}
This work has been performed in the framework of the H2020-ICT-2014-2 project 5G NORMA. The authors would like to acknowledge the contributions of their colleagues. This information reflects the consortium's view, but the consortium is not liable for any use that may be made of any of the information contained therein.

\bibliographystyle{IEEEtran}
\bibliography{references}

\end{document}